\documentclass[aps,print,showpacs,twocolumn]{revtex4}%
\usepackage{amsfonts}
\usepackage{amsmath}
\usepackage{amssymb}
\usepackage{graphicx}%
\begin{document}
\author{Gang Chen,$^{a,b\ }$Zidong Chen,$^{a}$ and J.-Q. Liang$^{b}$}
\affiliation{$^{a}$Department of Physics, Shaoxing College of Arts and Sciences, Shaoxing
312000, People's Republic of China}
\affiliation{$^{b}$Institute of Theoretical Physics, Shanxi University, Taiyuan 030006, China}
\title{{\LARGE Ground-State Properties for Coupled Bose-Einstein Condensates inside a
Cavity Quantum Electrodynamics}}

\begin{abstract}
We analytically investigate the ground-state properties of two-component
Bose-Einstein condensates with few $^{87}$Rb atoms inside a high-quality
cavity quantum electrodynamics. In the SU(2) representation for atom, this
quantum system can be realized a generalized Dicke model with a quadratic term
arising from the interatomic interactions, which can be controlled
experimentally by Feshbach resonance technique. Moreover, this weak
interspecies interaction can give rise to an important zero-temperature
quantum phase transition from the normal to the superradiant phases, where the
atomic ensemble in the normal phase is collectively unexcited while is
macroscopically excited with coherent radiations in the superradiant phase.
Finally, we propose to observe this predicted quantum phase transition by
measuring the direct and striking signatures of the photon field in terms of a
heterodyne detector out of the cavity.

\end{abstract}

\maketitle

Cavity quantum electrodynamics has an essential microwave photon field to
explore the important microscopic quantum phenomena in quantum optics since
the strong atom-field coupling strength, which dominates over the dissipative
losses of the quantum system, has been achieved\cite{1,2,3}. Recently, much
progress has been made in the observation of the motional dynamics\cite{4,5}
as well as the trapping\cite{6,7} and cooling\cite{8,9} of single atom within
the cavity mode. This also provides an interesting way to process quantum
information and implement quantum computing such as nonclassical light
sources\cite{10,11} and quantum state transfer\cite{12}.

Recently, Bose-Einstein condensate (BEC) coupled with a high-quality cavity
quantum electrodynamics has been attracted much attentions in both theory and
experiment. Despite BEC is a central goal for atom chips, it exhibits some
novel macroscopic quantum effects, which has never predicted and observed in
traditional atom. For example, atom-atom or atom-photon entanglement and
squeezing can be produced in the linear regime of collective atomic recoil
lasing, where the ground state of the condensate remains approximately
undepleted\cite{13,14,15}. By solving the three-mode master equation in the
Wigner representation, three-mode entanglement as well as two-mode atom-atom
and atom-radiation entanglement can be generally robust against losses and
decoherence, and thus regarded as a good candidate for the experimental
observation of entanglement in condensate systems\cite{16}. Moreover, a novel
cavity-mediated long-range atom-atom interactions has been also realized
theoretically\cite{17}. However, the experimental process has only been
considered recently \cite{18,19,20,21,22,23}. The main difficulties arise from
adverse vacuum requisites and sophisticated topological requirements on both
of these art technologies. Here we theoretically discuss two-component BECs
interacting with a high-quality cavity quantum electrodynamics. It has been
shown that in two-component BECs, the weak nonlinear interactions controlled
experimentally by Feshbach resonance technique\cite{24} can give rise to a
number of novel phenomena including instability\cite{25},
metastability\cite{26}, wave mixing\cite{27}, soliton\cite{28,29,30},
dynamical bifurcation\cite{31}, chaos\cite{32,33}, and Heisenberg-limited
Mach-Zahnder interferometry\cite{34}. Very recently, a novel second-order
quantum phase transition from the normal to the tunneling phases has been
predicted in the resonant case when two-component BECs are coupled with a
periodically driven laser field\cite{35}.

In this paper we mainly investigate the ground-state properties of
two-component BECs with few $^{87}$Rb atoms inside an ultrahigh finesse
optical cavity, which can support a single-mode photon. We show that in the
SU(2) representation for atom, this quantum system can be realized a
generalized Dicke model with a quadratic term arising from the mean-field
interatomic interactions. By means of Holstein-Primakoff transformation
\cite{36} and boson expansion method\cite{37}, the ground-state energy can be
approximately evaluated and show that the weak interspecies interaction can
lead to an important zero-temperature quantum phase transition from the normal
to the superradiant phases, where the atomic ensemble in the normal phase is
collectively unexcited while is macroscopically excited with coherent
radiations in the superradiant phase. Finally, we propose to observe this
predicted quantum phase transition by detecting the direct and striking
signatures of the photon field such as the well-measured intracavity intensity
$I\propto\left\vert \left\langle a\right\rangle \right\vert ^{2}$ in terms of
a heterodyne detector out of the cavity\cite{38}. Moreover, its essential
features of quantum criticality such as the scaling behavior, critical
exponent and universality are also given.

A physical realization shown in Fig.1 is that two condensates in different
hyperfine levels $\left\vert F=1,m_{f}=-1\right\rangle $ $(\left\vert
1\right\rangle )$ and $\left\vert F=2,m_{f}=1\right\rangle $ ($\left\vert
2\right\rangle $) are confined in a time-average, orbiting potential magnetic
trap. Usually, in the $^{87}$Rb system, an external laser is applied to induce
a Josephson-like coupling and the detuning of the laser is adiabatically
changed to produce various transitions between $\left\vert 1\right\rangle $
and $\left\vert 2\right\rangle $. However, here this tunable laser can be
displaced by the quantized field of the cavity mode after two condensates are
sent into the optical cavity. For a high-quality cavity with the length $178$
$\mu$m and the mode waist radius $26$ $\mu$m, the finesse is given by
$3\times10^{5}$ and the maximum coupling strength between the $^{87}$Rb atom
and the cavity field is $g=2\pi\times10.4$ MHz, which is larger than the
cavity field decay rate $\kappa=2\pi\times1.4$ MHz and the atom dipole decay
rate $\gamma=2\pi\times1.4$ MHz\cite{20,21,22}. Thus, in such a cavity the
single-mode photon can be considered and the quantum dissipation effect can be
also negligible. Finally, in the formalism of second quantization based on the
self-consistent Gross-Pitaevskii equations for the two-component BECs, the
total Hamiltonian for Fig.1 without rotating-wave approximation can be written as%

\begin{equation}
H=H_{ph}+H_{at}+H_{at-at}+H_{at-ph} \label{1}%
\end{equation}
with%

\begin{equation}
H_{ph}=\omega a^{+}a, \label{2}%
\end{equation}

\begin{align}
H_{at}  &  =%
%TCIMACRO{\dsum \limits_{l=1,2}}%
%BeginExpansion
{\displaystyle\sum\limits_{l=1,2}}
%EndExpansion
\int d^{3}\mathbf{r}\{\Psi_{l}^{+}(\mathbf{r})[-\frac{\hbar^{2}}{2m}%
\Delta+V_{l}(\mathbf{r})]\Psi_{l}(\mathbf{r})+\label{3}\\
&  \frac{q_{l}}{2}\Psi_{l}^{+}(\mathbf{r})\Psi_{l}^{+}(\mathbf{r})\Psi
_{l}(\mathbf{r})\Psi_{l}(\mathbf{r})\},\nonumber
\end{align}

\begin{equation}
H_{at-at}=\int d^{3}\mathbf{r}q_{1,2}\Psi_{1}^{+}(\mathbf{r})\Psi_{2}%
^{+}(\mathbf{r})\Psi_{1}(\mathbf{r})\Psi_{2}(\mathbf{r}),\label{4}%
\end{equation}

\begin{align}
H_{at-ph} &  =i\hbar\Omega\int d^{3}\mathbf{r}[\Psi_{1}^{+}(\mathbf{r}%
)\exp(i\mathbf{k}\cdot\mathbf{r})\Psi_{2}(\mathbf{r})+\label{5}\\
&  \Psi_{2}^{+}(\mathbf{r})\exp(-i\mathbf{k}\cdot\mathbf{r})\Psi
_{1}(\mathbf{r})](a-a^{+}),\nonumber
\end{align}
where $a$ and $a^{+}$ are the annihilation and creation operators of the
cavity mode with the frequency $\omega$; $\Psi_{l}(\mathbf{r})$ is the boson
field operator; $V_{l}(\mathbf{r})$ is the single magnetic trapped potential
with the frequencies $\omega_{i}(i=x,y,z)$; $q_{l}=\ 4\pi\hbar^{2}\rho_{l}/m$
is the intraspecies interactions with $\rho_{l}$ being the intraspecies
$s-$wave scattering length and $m$ being the atomic mass; $q_{1,2}=4\pi
\hbar^{2}\rho_{1,2}/m$ is the interspecies interactions with $\rho_{1,2}$
being the interspecies $s-$wave scattering length; $\Omega$ is the atom-cavity
coupling constant; and $\mathbf{k}$ is the wave vector of the quantized field
for the cavity mode.

It has been shown that the well-known two-mode approximation, which is defined
as $\Psi_{1}(\mathbf{r})=c_{1}\phi_{1}(\mathbf{r})$ and $\Psi_{2}%
(\mathbf{r})=c_{2}\phi_{2}(\mathbf{r})$ with $c_{1}$ and $c_{2}$ being the
annihilation boson operators, can be used to simplify the Hamiltonian (1) as
\begin{align}
H  &  =%
%TCIMACRO{\dsum \limits_{l=1,2}}%
%BeginExpansion
{\displaystyle\sum\limits_{l=1,2}}
%EndExpansion
(\omega_{l}c_{l}^{+}c_{l}+\frac{\eta_{l}}{2}c_{l}^{+}c_{l}^{+}c_{l}c_{l})+\chi
c_{1}^{+}c_{1}c_{2}^{+}c_{2}+\\
&  \lambda(c_{1}^{+}c_{2}+c_{2}^{+}c_{1})(a+a^{+})+\omega a^{+}a,\nonumber
\end{align}
where $\omega_{l}=\int d^{3}\mathbf{r}\{\phi_{l}^{\ast}(\mathbf{r}%
)[-\frac{\hbar^{2}}{2m}\Delta+V_{l}(\mathbf{r})]\phi_{l}(\mathbf{r})$,
$\eta_{l}=q_{l}\int d^{3}\mathbf{r}\left\vert \phi_{l}(\mathbf{r})\right\vert
^{4}$, $\chi=q_{1,2}\int d^{3}\mathbf{r}\left\vert \phi_{1}(\mathbf{r}%
)\right\vert ^{2}\left\vert \phi_{2}(\mathbf{r})\right\vert ^{2}$ and
$\lambda=i\hbar\Omega\int d^{3}\mathbf{r}\phi_{1}^{\ast}(\mathbf{r}%
)\exp(i\mathbf{k}\cdot\mathbf{r})\phi_{2}(\mathbf{r})=i\hbar\Omega\int
d^{3}\mathbf{r}\phi_{2}^{\ast}(\mathbf{r})\exp(-i\mathbf{k}\cdot
\mathbf{r})\phi_{1}(\mathbf{r})$ are the atom-cavity interacting constant and
assumed to be real for the sake of simplicity. However, it should be noticed
that this two-mode approximation can only be applied for a small number of
condensed atoms\cite{39,40}. A simple estimate shows that the number of atoms
$N$ should satisfy the condition such that $N\rho\leq r_{0}$, where $\rho$ is
a typical scattering length and $r_{0}$ is a measure of the trap size. If we
take into account large trap size $r_{0}=10$ $\mu m$ and the typical
scattering length $\rho=5$ nm\cite{41}, the two-mode approximation is valid
for $N\leq2000$.

The Hamiltonian (2) can be further simplified by applying the Schwinger
relations defined as $S_{x}=(c_{1}^{+}c_{2}+c_{2}^{+}c_{1})/2,\quad
S_{y}=(c_{1}^{+}c_{2}-c_{2}^{+}c_{1})/2i$, and$\quad S_{z}=(c_{1}^{+}%
c_{1}-c_{2}^{+}c_{2})/2$, where the Casimir invariant is $S^{2}=N(N/2+1)/2$.
The eigenvalues $m$ of operator $S_{z}$ represent the difference
$2(N_{1}-N_{2})$ in the number of atoms in different hyperfine levels, while
$S_{x}$ and $S_{y}$ take on the meaning of relative phase between two
condensates. Besides a trivial constant, the Hamiltonian (6) can be reduced to
a generalized Dicke Hamiltonian with a quadratic term%

\begin{equation}
H=qS_{z}^{2}+\omega_{0}S_{z}+2\lambda S_{x}(a^{+}+a)+\omega a^{+}a, \label{7}%
\end{equation}
where $q=[(\eta_{1}+\eta_{2})/2-\chi]$ and $\omega_{0}=\omega_{1}-\omega
_{2}+(N-1)(\eta_{2}-\eta_{1})/2\simeq N(\eta_{2}-\eta_{1})/2$ for the single
trapped potential. The Hamiltonian (7) describes the collective quantum
dynamics of two-component BECs with few $^{87}$Rb atoms inside an ultrahigh
finesse optical cavity. It can be seen clearly that the nonlinear quadratic
term arises from the mean-field interatomic interactions, which can be
controlled by the $s-$wave scattering lengths of atoms via Feshbach resonance
technique. For simplicity, here we assume that the weak nonlinear parameter
$q$ depends only on the interspecies interaction, namely, on the parameter
$\chi$ (or $\rho_{1,2}$) and consider the case of $\omega_{0}\geq0$, that is,
$\rho_{2}\geq\rho_{1}$. If the $s-$wave scattering lengths of atoms are chosen
as $\rho_{1,2}=(\rho_{1}+\rho_{2})/2$, the Hamiltonian (7) can be mapped into
the standard Dicke-like Hamiltonian $H_{D}=\omega a^{+}a+\omega_{0}%
S_{z}+2\lambda S_{x}(a^{+}+a)$\cite{42}.

In order to effectively describe the collective dynamical behavior of the
Hamiltonian (7), we should evaluate its ground-state energy. Since here the
trapped atom number is considered to be of order $10^{3}$, we can use the
mean-field approximation to arrive at the target. Following the procedure of
Ref.\cite{43}, we firstly employ the well-known Holstein-Primakoff
transformation, which is defined as $S_{+}=b^{+}\sqrt{N-b^{+}b}$, $S_{-}%
=\sqrt{N-b^{+}b}b$, and $S_{z}=(b^{+}b-N/2)$ with $[b,b^{+}]=1$\cite{36}, to
rewrite the Hamiltonian (7), apart from a trivial constant, as%

\begin{equation}
H=\omega a^{+}a+\tilde{\omega}_{0}(b^{+}b-N/2)+\frac{g}{\sqrt{N}}(b^{+}%
\sqrt{N-b^{+}b}+ \label{8}%
\end{equation}

\[
\sqrt{N-b^{+}b}b)(a^{+}+a)+\frac{\nu}{N}(b^{+}\sqrt{N-b^{+}b})(\sqrt{N-b^{+}%
b}b)
\]
where $\nu=-Nq$, $g=\sqrt{N}\lambda$ and $\tilde{\omega}_{0}=\omega
_{0}+q\simeq\omega_{0}$ since the parameter $q$ is a negligibly small number
compared with the parameter $\omega_{0}$. Secondly, we introduce the new boson
operators by setting $c^{+}=a^{+}+\sqrt{N}\alpha\ $and $d^{+}=b^{+}-\sqrt
{N}\beta$ to describe the collective behavior of the Hamiltonian (8). At last,
in terms of boson expansion method based on the introduced operators $c^{+}$
and $d^{+}$, the Hamiltonian (8) can be expanded by\cite{37}%

\begin{equation}
H=NH_{0}+N^{1/2}H_{1}+N^{0}H_{2}+\cdot\cdot\cdot\mathbf{\ } \label{9}%
\end{equation}
with $H_{0}=\omega\alpha^{2}+\omega_{0}(\beta^{2}-1/2)-4g\alpha\beta\sqrt
{k}+\nu\beta^{2}\sqrt{k}$, $H_{1}=(-\omega\alpha+2g\beta\sqrt{k}%
)(c^{+}+c)+[\omega_{0}\beta-(2g\alpha-\nu\beta\sqrt{k})(\sqrt{k}-\beta
^{2}/\sqrt{k})](d^{+}+d)$, $H_{2}=\omega c^{+}c+(\omega_{0}+4g\alpha
\beta/\sqrt{k}+\nu k-3\nu\beta^{2})d^{+}d+(g\alpha\beta/\sqrt{k}-\nu\beta
^{2})[(d^{+})^{2}+d^{2}]+(g\alpha\beta^{3}/2k^{3/2})(d^{+}+d)^{2}+(g\sqrt
{k}-\beta^{2}/\sqrt{k})(c^{+}+c)(d^{+}+d)$, where $k=1-\beta^{2}$.

The critical transition point can be derived from the condition $H_{1}\equiv
0$, which means that\ for any quantum system the free energy should be fixed
at the minimum value. Therefore, two different cases for $\delta>1$ and
$\delta<1$ with $\delta=\omega\omega_{0}/N(4\lambda^{2}+\omega q)$ must be
considered. The corresponding auxiliary parameters $\alpha$\ and $\beta$\ are
given by%

\begin{equation}
\alpha=\left\{
\begin{array}
[c]{c}%
0,\text{ }\delta>1\\
\frac{\sqrt{N}\lambda\sqrt{1-\delta^{2}}}{\omega},\delta<1
\end{array}
\right.  ,\beta=\left\{
\begin{array}
[c]{c}%
0,\text{ }\delta>1\\
\sqrt{\frac{(1-\delta)}{2}},\delta<1
\end{array}
\right.  \label{10}%
\end{equation}
and the interesting critical transition point is given by $\delta=$ $1$, namely,%

\begin{equation}
q_{c}=\frac{\omega_{0}}{N}-\frac{4\lambda^{2}}{\omega}. \label{11}%
\end{equation}
Since the many-body interactions in the two-mode approximation produce only
small modification of the ground-state properties, the wavefunctions of the
macroscopic condensate states for the single magnetic trap can be given, if
the acting role of the gravity is omitted, by $\phi_{l}(\mathbf{r})=\pi
^{-3/4}(d_{x}d_{y}d_{z})^{-1/2}\exp[-(x^{2}/d_{x}^{2}+y^{2}/d_{y}^{2}%
+z^{2}/d_{z}^{2})/2]$ with $d_{x}=\sqrt{\hbar/m\omega_{x}}$, $d_{y}%
=\sqrt{\hbar/m\omega_{y}}$ and $d_{z}=\sqrt{\hbar/m\omega_{z}}$. When the
relevant parameters are chosen as $\omega_{x}=\omega_{y}=2\pi\times290$ Hz,
$\omega_{z}=2\pi\times450$ Hz, $m=1.45\times10^{-25}$ kg, $\rho_{1}=3.70$ nm,
$\rho_{2}=5.70$ nm, and $N=1000$, respectively, the parameter $\omega_{0}$ in
the unit of $\hbar$ can be approximately calculated as $\omega_{0}\simeq
N\hbar^{2}(\rho_{2}-\rho_{1})/\sqrt{2\pi}d_{x}d_{y}d_{z}m\simeq2866$ Hz. If
the frequency of the photon and the atom-field coupling strength are
considered by $\omega=4\times10^{8}$ MHz and $\lambda=2\pi\times5$
MHz\cite{20,21,22}, respectively, the critical transition point is evaluated
as $q_{c}=-7$ Hz, whose corresponding interspecies $s-$wave scattering length
is given by $(\rho_{1,2})_{c}=7.14$ nm. It seems that this critical parameter
$(\rho_{1,2})_{c}$ can be achieved by current Feshbach resonance technique,
which awaits experimental validation.

In the case of $\rho_{1,2}>(\rho_{1,2})_{c}$ ($\delta>1$), the ground-state
energy is given by $(E_{0})_{nor}=-N\omega_{0}/2$, which depends only on the
parameter of the BECs. However, in the case of $\rho_{1,2}<(\rho_{1,2})_{c}$
($\delta<1$), the ground-state energy becomes $(E_{0})_{\sup}=-N[N(\lambda
^{2}/\omega+q/4)(1-\delta^{2})+\omega_{0}\delta/2]$, which is dependent of the
parameters of both the BECs and the cavity field. The physics may be
understood as follows. For the case $\rho_{1,2}>(\rho_{1,2})_{c}$, the quadric
term $(qS_{z}^{2})$ in the Hamiltonian (7) plays an important role, which
means that the transition between two condensates is suppressed and therefore
no collective excitation happens. However, for the other case $\rho
_{1,2}<(\rho_{1,2})_{c}$, the energy of the term $S_{x}(a^{+}+a)$ in the
Hamiltonian (7) dominates, which implies that both the internal Josephson
tunneling and the macroscopic collective excitation occur. Therefore, we can
call the normal phase when $\rho_{1,2}>(\rho_{1,2})_{c}$ and the superradiant
phase when $\rho_{1,2}<(\rho_{1,2})_{c}$. Fig.2 shows the scaled ground-state
energy $E_{0}/N$ and its second-order derivative with respect to the
intraspecies $s-$wave scattering length $\rho_{1,2}$ as a function of
$\rho_{1,2}$. It can be seen that the second-order derivative of $E_{0}/N$
possesses a discontinuity at the transition point $(\rho_{1,2})_{c}$, which
clearly illustrates the nature of second-order phase transition (the
first-order derivative of $E_{0}/N$ is continuous at the transition point).
Fig.3 shows the scaled ground-state atom population between two condensates%

\begin{equation}
\frac{\Delta N}{N}=2\left\langle S_{z}\right\rangle =\left\{
\begin{array}
[c]{l}%
-\frac{\omega\omega_{0}}{N(4\lambda^{2}+\omega q)}\text{, }\rho_{1,2}%
<(\rho_{1,2})_{c}\\
-1\text{,}\ \rho_{1,2}>(\rho_{1,2})_{c}%
\end{array}
\right.  \label{12}%
\end{equation}
and its first-order derivative with respect to the intraspecies $s-$wave
scattering length $\rho_{1,2}$ as a function of $\rho_{1,2}$, which also
demonstrates the superradiant phase transition.

In the rest part of this paper we discuss how to observe this predicted phase
transition in current experimental setups. In general, the many-body quantum
pseudospin state in this quantum system is not accessible to observe quantum
phase transition. However, here we propose to detect the direct and striking
signatures of the photon field such as the well-measured intracavity intensity
$I\propto\left\vert \left\langle a\right\rangle \right\vert ^{2}$ in terms of
a heterodyne detector out of the cavity\cite{38}. In terms of the auxiliary
parameters $\alpha$ given by Eq.(10), the scaled ground-state intracavity
intensity is evaluated as%

\begin{equation}
\frac{I}{N}\propto\left\{
\begin{array}
[c]{l}%
\frac{\lambda^{2}[N^{2}(4\lambda^{2}+\omega q)^{2}-\omega^{2}\omega_{0}^{2}%
]}{N\omega^{2}(4\lambda^{2}+\omega q)^{2}}\text{, }\rho_{1,2}<(\rho_{1,2}%
)_{c}\\
0\text{,}\ \rho_{1,2}>(\rho_{1,2})_{c}%
\end{array}
\right.  \label{13}%
\end{equation}
Fig.4 shows the scaled ground-state intracavity intensity $I/N$ and its
first-order derivative with respect to the intraspecies $s-$wave scattering
length $\rho_{1,2}$ as a function of $\rho_{1,2}$. It is interesting that this
predicted quantum phase transition characterized by the non-analyticity of the
scaled ground-state intracavity intensity $I/N$ is remarkably of the
first-order. When the intraspecies $s-$wave scattering length $\rho_{1,2}$
approaches the critical value $(\rho_{1,2})_{c}$, the scaled ground-state
intracavity intensity $I/N$ vanishes as%

\begin{equation}
\frac{I}{N}[\rho_{1,2}\rightarrow(\rho_{1,2})_{c}]\sim\left\vert \rho
_{1,2}-(\rho_{1,2})_{c}\right\vert . \label{14}%
\end{equation}
Since the diverging characteristic length scale is $\zeta\sim\left\vert
\rho_{1,2}-(\rho_{1,2})_{c}\right\vert ^{-v}$ with $v=1/2$, the critical
exponent for the scaled ground-state intracavity intensity can be derived from
$I/N\sim\left\vert \rho_{1,2}-(\rho_{1,2})_{c}\right\vert ^{zv}$ by $z=2$,
which shows the universality principle of quantum phase transition\cite{44}.

In conclusion, we have analytically discussed the ground-state properties of
two-component BECs of $^{87}$Rb atoms inside an ultrahigh finesse optical
cavity supporting a single-mode photon. It has been shown that the weak
interspecies mean-field interaction can give rise to a novel zero-temperature
quantum phase transition from the normal to the superradiant phases, which
means that the weak microscopic nonlinear interaction can lead to a strong
behavior in macroscopic scale. Moreover, we have proposed to observe this
predicted quantum phase transition by measuring the direct and striking
signatures of the photon field such as the well-measured intracavity intensity
$I\propto\left\vert \left\langle a\right\rangle \right\vert ^{2}$ in terms of
a heterodyne detector out of the cavity. It is also interesting that this
scaled ground-state intracavity intensity $I/N$ is remarkably of the
first-order at the transition point $(\rho_{1,2})_{c}$.

We thank Prof. S. L. Zhu for helpful discussions and valuable suggestions.
This work was supported by the Natural Science Foundation of China under Grant
No.10475053 and by the Natural Science Foundation of Zhejiang Province under
Grant No.Y605037.

\end{document}